\documentclass[aps,prl,showpacs,twocolumn,amssymb,groupedaddress]{revtex4}

\usepackage{times}      
\usepackage{graphicx}   
\usepackage{epsfig}      
\usepackage{dcolumn}    

\begin{document}
\title{Challenging the paradigm of singularity excision in gravitational collapse}

\author{Luca~Baiotti$^{1}$, Luciano~Rezzolla$^{1,2,3}$}

\affiliation{$^1$Max-Planck-Institut f\"ur Gravitationsphysik,
        Albert-Einstein-Institut, 14476 Golm, Germany}
\affiliation{$^2$SISSA, International School for
        Advanced Studies and INFN, Via Beirut 2, 34014 Trieste, Italy}
\affiliation{$^3$Department of Physics, Louisiana State University, Baton
        Rouge, LA 70803 USA}

\date{\today}

\begin{abstract} 
	A paradigm deeply rooted in modern numerical relativity
        calculations prescribes the removal of those regions of the
        computational domain where a physical singularity may develop. We
        here challenge this paradigm by performing three-dimensional
        simulations of the collapse of uniformly rotating stars to black
        holes without excision. We show that this choice, combined with
        suitable gauge conditions and the use of minute numerical
        dissipation, improves dramatically the long-term stability of the
        evolutions.  In turn, this allows for the calculation of the
        waveforms well beyond what previously possible, providing
        information on the black-hole ringing and setting a new mark on
        the present knowledge of the gravitational-wave emission from the
        stellar collapse to a rotating black hole.
\end{abstract}



\maketitle



Numerical relativity simulations have recently recorded important
breakthroughs, which have allowed for long-term stable and accurate
calculations of curved and highly dynamical spacetimes, thus increasing
their potential of providing physically relevant predictions for
gravitational-wave astronomy. Behind this rapid progress are various
novel approaches, some of which involve the dropping of assumptions or
techniques that were considered to be important or simply necessary.
Firstly, the dismissal of the $3+1$ ADM formulation of the field
equations, which, after large-scale efforts~\cite{Cook97a}, has shown not
to have the stability properties necessary for long-term fully
three-dimensional (3D) simulations. In lieu of the ADM equations, new
formulations have been either reconsidered (as in the case of the
conformal and traceless formulation of the ADM
equations~\cite{Nakamura87}) or investigated for the first time in 3D
simulations (as in the case of the harmonic formulation of the Einstein
equations~\cite{Pretorius:2005gq}). Both approaches have been shown to
provide long-term stability on timescales sufficiently large to evolve
accurately a large class of spacetimes, including black
holes~\cite{Pretorius:2005gq, Baker05a, Campanelli:2005dd} and neutron
stars~\cite{Shibata:2003iw,Duez:2002bn}. Secondly, the abandoning of the
use of numerical grids with uniform spacing, in lieu of which several
codes now use mesh-refinement techniques (either
fixed~\cite{Schnetter-etal-03b} or with adaptivity~\cite{Baker05a}).
This conceptually simple but technologically challenging improvement
allows to concentrate computational resources where the truncation error
needs to be the smallest, while saving them where they may not be
needed. In turn, non-uniform grids have allowed to place the outer
boundaries of the computational domain at very large distances, thus
reducing the influence of inaccurate outer-boundary conditions and making
it possible to have the wave-zone within the domain and to extract there
the precious gravitational-wave signal~\cite{Baiotti04b}.  Thirdly, for
some years now, successful long-term 3D evolutions of black-hole vacuum
spacetimes have been possible only thanks to the use of excision
techniques (see, {\it e.g.},~\cite{Seidel92a} for a first example), that
is by ignoring the spacetime regions inside black-hole horizons. These
are causally disconnected from the outside and should have no effect on
the rest of the evolution as long as a suitable treatment of the
equations is made at the excision surface. Furthermore, recent
simulations of the collapse of rotating neutron stars to Kerr black
holes~\cite{Baiotti04,Duez:2002bn} have shown the effectiveness of
excision techniques also for spacetimes with matter, where they were
applied separately to the field equations and to the hydrodynamical
equations. In those simulations, in fact, the use of excision has
extended considerably the lifetime of the simulations, allowing for an
accurate investigation of the dynamics of the trapped surfaces formed
during the collapse and for the extraction of the first gravitational
waveforms from 3D collapse to rotating black holes~\cite{Baiotti04b}.

Although the assumption that a region of spacetime that is causally
disconnected can be ignored without this affecting the solution in the
remaining portion of the spacetime is certainly true for signals and
perturbations travelling at physical speeds, numerical signals, such as
gauge waves or constraint violations, may travel at velocities larger
than that of light and thus leave the physically disconnected
region. Indeed, this is what is commonly observed when excising a
topologically spherical surface in Cartesian coordinates within a conformal
and traceless formulation of the Einstein equations and without the
massive use of numerical dissipation as in ref.~\cite{Pretorius:2005gq}.

In this paper we challenge the paradigm of singularity excision and show
that accurate numerical simulations can be carried out even in the
absence of an excised region and independently of whether the spacetime
is vacuum and the singularity modelled as a ``puncture''~\cite{Baker05a,
Campanelli:2005dd}. In addition, we show that the absence of an excised
region improves dramatically the long-term stability, allowing for the
calculation of the gravitational waveforms well beyond what previously
possible and past the black-hole quasi-normal-mode (QNM) ringing.

The simplest and most impressive way of proving the improvements
resulting from not excising the region inside the apparent horizon (AH)
is by reconsidering the same initial data of ref.~\cite{Baiotti04b},
which lead to gravitational collapse to rotating black holes. More
specifically, we consider rotating relativistic stars calculated as
equilibrium solutions of the Einstein equations, with a polytropic
equation of state $p=K \rho^{\Gamma}$, with $\Gamma=2$ and with the
polytropic constant initially set to $K_{_{\rm ID}}=100$~\footnote{This
is subsequently reduced of 2\% to trigger the collapse.}. To illustrate
the collapse for the range of neutron stars rotating either
very slowly or near the mass-shedding limit, we consider two
representative models, indicated as D1 and D4 in ref.~\cite{Baiotti04},
with different initial angular momenta. The first one is a slowly
rotating star of mass $M=1.67\,M_{\odot}$, circumferential equatorial
radius $R_e=11.43$ km and $a \equiv J/M^2 =0.21$, where $J$ is the
angular momentum; the second one has with
$M=1.86\,M_{\odot}$, $R_e=14.25$ km and is rotating close to the
mass-shedding limit with $a=0.54$.

The numerical methods used for these evolutions are the same as discussed
in ref.~\cite{Baiotti04b}. In particular, we use a conformal and
traceless formulation of the ADM equations~\cite{Alcubierre99d}, which
are solved with fourth-order finite-difference operators, while we evolve
the hydrodynamical equations with the {\tt Whisky}
code~\cite{Baiotti03a}, implementing high-resolution shock-capturing
(HRSC) techniques with a variety of approximate Riemann solvers and
reconstruction algorithms and an overall second-order truncation
error~\cite{Baiotti04}. The numerical grid setup makes use of the same
mesh refinement implemented in the numerical infrastructure described in
ref.~\cite{Schnetter-etal-03b}. Besides a fixed number of refinement
levels which are present already on the initial slice, new refined levels
are added at predefined positions during the evolution. More
specifically, as the star collapses, the innermost (most refined)
grid-box is progressively further refined with box-in-box grids, the
final one having a resolution of $\Delta x \simeq 0.02\,M$; the outermost
grid, instead, which is not further refined, has a resolution of $\Delta
x \simeq 1.5\,M$, sufficient to capture the details of the gravitational
radiation. In this way, our outer boundaries are placed at $\simeq 165\,M
$, so as to minimize the influence of our imperfect outer-boundary
conditions on the very small gravitational-wave signal. We note that the
``switching-on'' of different levels of resolution 
does have a small but appreciable effect on its dynamics, essentially due
to the necessary interpolation to the new refined grids, which also
changes the growth-time of the most unstable, exponentially growing
mode. This effect is negligible in the evolution of the matter but
can influence the waveforms. To provide a direct comparison with~\cite{Baiotti04b}, 
we have here followed the same approach.

\begin{figure}
\centerline{
\psfig{file=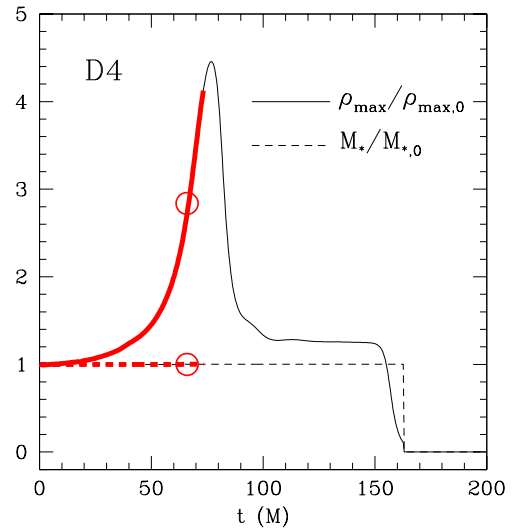,angle=-0,width=7.5cm}
}
\vspace{-0.25cm}
\caption{Evolution of the normalized maximum of the rest-mass density
  $\rho_{\rm max}$ (continuous line) and of the total rest-mass $M_*$
  (dashed line) for model D4. Circles indicate the times at
  which the AH is first found and thick lines the
  results obtained with excision.}
\vspace{-0.5cm}
\label{matter_dynamics}
\end{figure}

Fig.~\ref{matter_dynamics} summarises the dynamics of the matter for
model D4 by showing the time evolution of the maximum of the rest-mass
density $\rho_{\rm max}$ (continuous lines) and of the total rest-mass
$M_*$ (dashed lines) when normalized to their initial values (a similar
behaviour is observed also for the slowly rotating model D1). The
evolution obtained without excision is to be contrasted with the one in
which the excision is made (the used excision technique is described in detail in
ref.~\cite{Alcubierre:2004bm}) and which is indicated with thick
lines. In that case, the excision was started soon after an
AH~\cite{Thornburg2003:AH-finding} was found (this is marked with circles
on the two curves) and the corresponding evolution terminated at $t
\gtrsim 77\,M$, when the code crashed with large violations in the
Hamiltonian constraint.  Note that while the maximum values for the
rest-mass density attained during the evolution are comparable in the two
cases, the evolution without excision can be carried out to much later
times ({\it i.e.} $t \gtrsim 300\,M$) without appreciable loss of
accuracy or sign of instability. Indeed, as the matter collapses and
concentrates over a very few gridpoints (and ultimately on only one), the
high accuracy of the HRSC methods is able to conserve the rest mass to
very high precision with a loss of less than 0.2\% up to when the
rest-mass density distribution is diffused as a result of
the poorly resolved gradients.

While very little extra is needed for the evolution of the
hydrodynamical quantities in the absence of an excision algorithm, a
stable evolution of the Einstein equations requires at least two
important ingredients. The first one is represented by gauge
conditions for the lapse function $\alpha$ with suitable
singularity-avoidance properties. Our experience has shown that
hyperbolic $K$-driver slicing conditions of the form $\partial_t
\alpha = - f(\alpha) \; \alpha^2 (K-K_0)$ , with $f>0$ and $K_0$ being
the trace of the extrinsic curvature at $t=0$, are essential to
``freeze'' the evolution in those regions of the computational domain
inside the AH, where the metric functions experience the growth of
very large gradients. In practice, we confirm that using the
generalized ``$1+$log'' slicing condition~\cite{Bona94b} as obtained
by setting $f=2/\alpha$ provides the desired singularity avoidance and
is computationally efficient. With a good choice
for the slicing condition, the results do not depend sensitively on
the gauge condition for the shift. We have found that the use of the
``Gamma-driver'' shift conditions discussed in ref.~\cite{Baiotti04}
is sufficient to compensate for the ``slice-stretching'' induced by
the singularity-avoiding slicing.

The second important ingredient is the introduction of an artificial
dissipation of the Kreiss-Oliger type~\cite{Kreiss73} on the
right-hand-sides of the evolution equations for the spacetime variables
and the gauge quantities (No dissipation is introduced for the
hydrodynamical variables.). This is needed mostly because all
the field variables develop very steep gradients in the region inside the
AH. Under these conditions, small high-frequency oscillations (either
produced by finite-differencing errors or by small reflections across the
refinement boundaries) can easily be amplified, leave the region inside
the AH and rapidly destroy the solution. In practice, for any
time-evolved quantity $u$, the right-hand-side of the corresponding
evolution equation is modified with the introduction of a term of the
type ${\cal L}_{\mbox{\tiny diss}}(u) = -\varepsilon \Delta x_i^3
\partial^4_{x_i} u$, where $\varepsilon$ is the dissipation
coefficient, which is allowed to vary in space. We have
experimented with configurations in which the coefficient was either
constant over the whole domain or larger for the gridpoints inside the
AH. No significant difference has been noticed between the two
cases. Much more sensitive is instead the choice of the value of
$\varepsilon$.  In the simulations reported here we have set $\varepsilon
= 0.01$ for model D1 and $\varepsilon = 0.0075$ for D4,
respectively. However, smaller values ({\it e.g.}  $\varepsilon = 0.005$)
are not sufficient to yield the long-term stability discussed here, while
larger values ({\it e.g.}  $\varepsilon = 0.05$ and $\varepsilon = 0.01$,
respectively) alter significantly the waveforms, which would not match
the ones obtained without dissipation up until the the latter can be computed.

\begin{figure}
\centerline{
\psfig{file=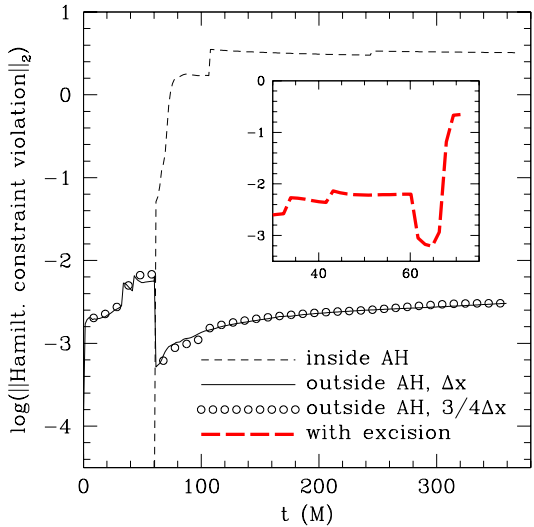,angle=-0,width=7.5cm}
}
\vspace{-0.25cm}
\caption{Evolution of the $L2$ norm of the Hamiltonian-constraint
  violation for model D1. Different curves refer to the violation
  computed only inside the AH (short-dashed line), only outside it
  (continuous line) or when the excision is made (long-dashed line in the
  inset). The circles show the rescaled violation for a higher
  resolution.}
\vspace{-0.5cm}
\label{ham_norm2_compare}
\end{figure}

In the region within the AH, as a consequence of large gradients in the spacetime variables, which
cannot be resolved sufficiently despite the use of several mesh-refinement levels, the solution of
the Einstein equations becomes increasingly inaccurate as the collapse
proceeds. Fig.~\ref{ham_norm2_compare} offers a measure of this loss of accuracy by showing the time
evolution of the $L2$ norm of the Hamiltonian-constraint violation for model D1. To distinguish the
different amplitudes of the errors, the violation has been computed separately for the domain inside
the AH and for the rest of the grid.

\begin{figure*}
\centerline{
\psfig{file=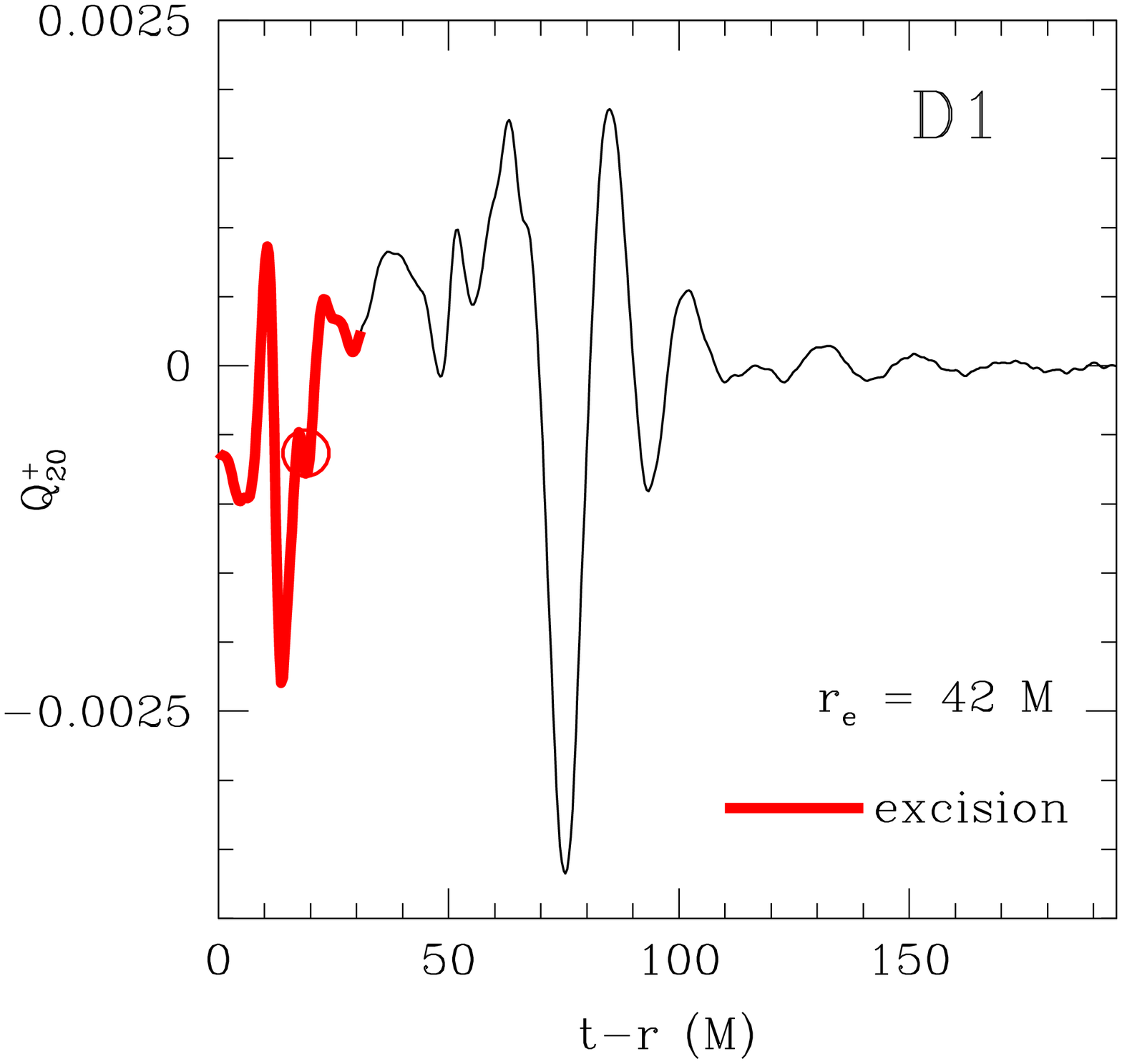,angle=-0,width=7.5cm}
\hspace{0.5cm}
\psfig{file=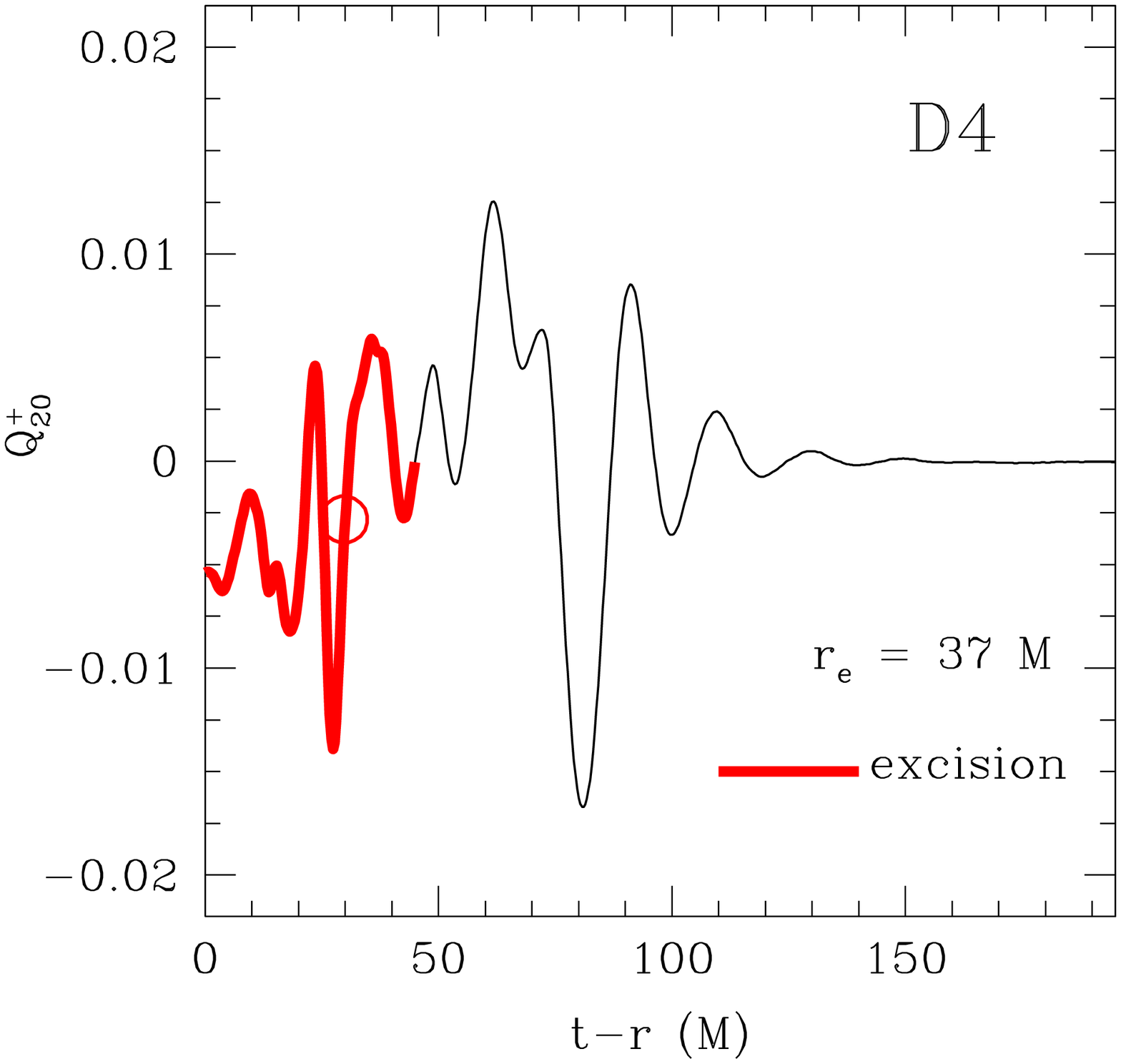,angle=-0,width=7.5cm}
        }
\vspace{-0.25cm}
\caption{Lowest gauge-invariant multipole for a slowly rotating star
  (left panel) and a rapidly rotating one (right panel). Thick lines
  refer to evolutions carried out with excision and terminate at a code
  crash, while the circles indicate the time of the AH formation time.}
\vspace{-0.5cm}
\label{waveforms_at_70}
\end{figure*}

Several aspects of Fig.~\ref{ham_norm2_compare} are worth
noting. First, the two errors are considerably different in amplitude,
the one inside the AH being much larger. Second, the sudden decrease in the 
long-dashed and continuous lines at the time of AH formation at 
$t\simeq 60\,M$ is due to the fact that some gridpoints (those with the 
largest violations) are no longer included in the
calculation of the $L2$ norm. Third, the violation outside
the AH grows only at a slow pace over the remaining evolution. Finally,
shown with circles is the violation for a high-resolution
simulation. The values are rescaled to second order before AH formation
(when the dominant truncation error comes from hydrodynamics and the HRSC
methods used are only second-order) and to third order afterwards (when
the smooth accretion flow of the atmosphere boosts the accuracy of the
hydrodynamics to third-order). No convergence of the violation
is seen inside the AH. Overall, Fig.~\ref{ham_norm2_compare} shows that
the solution of the Einstein equations inside the AH is by and large
incorrect, but these errors remain confined within the AH and do not
contaminate the overall accuracy of the simulation. Shown for comparison
in the inset is the violation computed when the singularity
is excised, showing an exponential growth soon after the AH is found.

Note that the ability to perform these long-term evolutions cannot be
related in any manner to the use of a ``puncture'' prescription for the
physical singularity, as recently done in simulations of binary black
holes~\cite{Baker05a,Campanelli:2005dd}. However, it is possible that the
stability provided here by the singularity-avoiding gauge and made
numerically more robust by the use of dissipation is closely related to
the one discussed for punctures~\cite{Hannametal06}.

Besides a long-term stability and the possibility of following the
collapse well beyond what was possible with the use of excision
techniques, the most dramatic advantage produced by the approach
suggested here can be appreciated in the study of the gravitational
radiation produced during the collapse. As in ref.~\cite{Baiotti04b}, we
have extracted the gravitational-wave information through an approach in
which the spacetime is matched with the non-spherical perturbations of a
Schwarzschild black hole described in terms of gauge-invariant odd
$Q^{({\rm o})}_{\ell m}$ and even-parity $\Psi^{({\rm e})}_{\ell m}$
metric perturbations, where $\ell, m$ are the indices of the angular
decomposition.  In Fig.~\ref{waveforms_at_70} we report the lowest-order
multipoles $Q^{+}_{20} \equiv \lambda\Psi^{({\rm e})}_{20}$, where
$\lambda \equiv \sqrt{{2(\ell+2)!} / {(\ell-2)!}}$.  The left panel, in
particular, shows the signal detected by an observer at a distance of
$42\,M$ for the slowly rotating model D1; the right panel, instead, shows
the same multipole extracted at $37\,M$ for the rapidly rotating model
D4. In both cases the thick lines refer to the evolutions carried out
with the excision technique and terminate when the code crashed ({\it
cf.} left panel of Fig.~1 in ref.~\cite{Baiotti04b}).  This comparison
shows that it is now possible to detect the complete gravitational-wave
signal produced by the collapse of a relativistic star to a rotating
black hole well beyond what was previously possible in either
2D~\cite{Stark85} or 3D simulations~\cite{Baiotti04b}. The estimated
error on the phase is $\lesssim 1\%$ and $\lesssim 5\%$ on the L2 norm of
the amplitude.

Fig.~\ref{waveforms_at_70} also highlights that the signal can be rather
different both in amplitude and form in the two cases with the exception
of the final parts, when the signal is dominated by the black-hole QNM
ringing.  This happens between $\simeq 80\,M$ and $\simeq 120\,M$, where
the signal we extract matches very well with a perturbative one computed
using the frequencies given in ref.~\cite{Berti06c} for the angular
momenta and masses of our models. The very good agreement with the
perturbative results is an important confirmation of the accuracy of our
results, which are intrinsically plagued by the extreme
weakness of the emitted gravitational radiation. Furthermore, the
richness of details in the two waveforms opens the prospects that a
careful characterization of the waveforms will provide important
information on the properties of the black hole as well as on those of
the collapsing matter~\cite{Baiottietal06}.

A straightforward analysis of the {\it now-complete} gravitational-wave
signal computed in these simulations allows to improve the estimates
provided in ref.~\cite{Baiotti04b} for the detectability of the
gravitational collapse of a uniformly rotating polytropic star at a
distance of 10 kpc. More specifically, the energy
efficiency in the emission of gravitational radiation is $E_{_{\rm D1}}/M = 3.3 \times 10^{-7}$
and $E_{_{\rm D4}}/M = 3.7 \times 10^{-6}$ respectively, with an overall accuracy of $\sim
10\%$. The resulting signal-to-noise ratios are
then: $(S/N)_{_{\rm D1-D4}}^{\rm Virgo} \simeq 0.27-2.1$, $(S/N)_{_{\rm
D1-D4}}^{^{\rm advLIGO}} \simeq 1.2-11$, and $(S/N)_{_{\rm D1-D4}}^{^{\rm
Dual}} \simeq 3.3-28$ for detectors such as Virgo/LIGO, advanced LIGO or
Dual~\cite{Bonaldi:2003}.

In conclusion, we have presented the first 3D calculations of the
gravitational collapse of uniformly rotating stars to black holes without
excision. This choice, together with suitable gauge conditions and the
use of minute numerical dissipation, improves dramatically the long-term
stability of the evolutions, providing the most accurate waveforms of
this process to date. While our approach represents a challenge to the
paradigm of singularity excision, it does not necessarily imply that all
excision techniques should be expected to lead to instabilities. Rather,
it highlights that, for a conformal traceless formulation of the Einstein
equations and in highly dynamical spacetimes, the excision of a spherical
surface in Cartesian coordinates may be more of a problem than a
solution.

\medskip
\noindent We thank D. Pollney, B. Szil\'agyi and J. Thornburg for useful
discussions. Support comes also through the SFB-TR7 of
the German DFG and the Italian INFN ``OG51''.


\end{document}